\documentclass[12pt]{article}
\usepackage{graphicx}
\usepackage[utf8]{inputenc}
\usepackage[english
]{babel}
\def\baselinestretch{1.5}
\begin{document}
 \begin{center}
	\bf{Probability representation of photon states and tomography}\\
\end{center}
\bigskip

\begin{center} {\bf V. N. Chernega$^1$, O. V. Man'ko$^{1,2}$, V. I. Man'ko$^{1,3,4}$}
\end{center}

\medskip

\begin{center}
	$^1$ - {\it Lebedev Physical Institute, Russian Academy of Sciences\\
		Leninskii Prospect 53, Moscow 119991, Russia}\\
	$^2$ - {\it Bauman Moscow State Technical University\\
		The 2nd Baumanskaya Str. 5, Moscow 105005, Russia}\\
	$^3$ - {\it Moscow Institute of Physics and Technology (State University)\\
		Institutskii per. 9, Dolgoprudnyi, Moscow Region 141700, Russia}\\
	$^4$ - {\it Russian Quantum Center\\
		Skolkovo, Moscow 143025, Russia}
	
	Corresponding author e-mail: mankovi@lebedev.ru
\end{center}

	\section*{Abstract}
	We give a review of the tomographic probability representation of quantum mechanics. We present the formalism of quantum states and quantum observables using the formalism of standard probability distributions and classical-like random variables. We study the coherent and number states of photons in the probability representation and obtain the evolution equation and energy spectra in the form of equations for probability distributions.
	
	\section{Introduction}
	The formalism of quantum mechanics is connected with the notion of Hilbert space vector and density operators and corresponding wave functions and density matrices  associated with quantum system states. In this formalism, quantum observables are identified with Hermitian operators and Hermitian matrices. There exist different representations of the operators acting in the Hilbert space of quantum states like the Wigner function, Husimi--Kano functions, and  Glauber--Sudarshan quasidistributions. Discrete analogs of the phase--space quasidistributions were also considered for Pauli spin states; see, e.g., \cite{Stratonovich}. Some aspects of associating quantum states with probability distributions were discussed in \cite{Mielnik,Wooters}. A detailed suggestion to identify quantum states with standard probability distributions was performed  in \cite{ManciniTombesiPLA,ManciniTombesiFoundPhys}. This suggestion is related to the technique of quantum tomography of photon states \cite{RaymerPRL93}. In the tomographic approach, the photon states are identified with the Wigner quasidistribution. This function is reconstructed from the measured optical tomogram, which is the probability distribution of the photon quadrature. Using the integral relation of the tomogram to the Wigner function be means of the Radon transform, the photon state identified with the quasidistribution is measured. In \cite{ManciniTombesiPLA,ManciniTombesiFoundPhys} it was suggested to consider the tomographic probability distribution not as technical tool to reconstruct the Wigner function of the state but to consider this probability distribution as a primary notion of the quantum state for all systems. This approach was studied in \cite{Man1JRLR}. In this connection, the quantum evolution equations equivalent to Schrodinger equation and von Neumann equation for the Wigner function were found \cite{AmosovKorennoyMankoPhysRev}. It is worth adding that the Pauli equation for the tomographic probability distribution is discussed in \cite{Pualeq}. 
	The relation of the tomographic picture of the states in quantum and classical mechanics was elucidated in \cite{Man1JRLR}. The probability representation of quantum states was used to investigate the Schr\"odinger cat paradox and the EPR-paradox in \cite{Foukzon}. The quantum-to-classical limits for quantum tomograms were studied and compared with
	the corresponding classical tomograms in \cite{Marmo3}. 
	Different problems of quantum--classical 
	phenomena and their connections with probability representation were investigated in \cite{KhrennikobAlo,Gousson,Belousov,Bazrafkan,Stornaiolo,Facci2,VALLONE}. The review and development of quantum tomography and its applications, as well as the review of the probability representation of quantum mechanics were given in \cite{MankoPhysScr2015150,IbortMarmoPhysSCR150,OlgaConfSer,Ibort,MarmoSudarshanAsoreyPRA}. Examples of probability representation of spin-1/2 states were given, e. g. in  \cite{MarmoVitaleJPhysA,Avanesovprep2019,AvanesovManko}. The relation of the approach to star--product formalism was considered in \cite{OlgaMarmo2001,OlgaMarmo2002,Patrizia}.
	
	In this work, we consider the examples of qubit and qudit as well as photon states and observables in the probability representation of quantum mechanics. This  paper is organized as follows. In Sec. 2, we study the qubit (spin-1/2) state. In Sec. 3, we consider the harmonic-oscillator coherent states and other oscillator states in the probability representation. Conclusions and prospectives are presented in Section 4.

	\section{Qubit states in probability representation}
	The qubit state density $2\times2$-matrix is determined by three parameters. This matrix 
	\begin{equation}\label{eq.1.1}
	\rho=
	\left(\begin{array}{cc}
	\rho_{11}&\rho_{12}\\
	\rho_{21}&\rho_{22}\end{array}\right),
	\end{equation}
	satisfies the Hermiticity conditions $\rho^\dag=\rho$, and its trace $\mbox{Tr}\rho=1$. The nonnegativity condition of the matrix eigenvalues provides the inequality 
	\begin{equation}\label{eq.1.2}
	\rho_{11}\rho_{22}-|\rho_{12}|^2\geq0.
	\end{equation}
	There exist different possibilities to  parametrize this matrix. One of the possibilities is to introduce Bloch ball parameters $-1\leq x,y,z\leq1$. One has $\rho_{11}=\frac{1+z}{2},\,\rho_{12}=\frac{x-iy}{2}$. These parameters satisfy the inequality
	\begin{equation}\label{eq.1.3}
	x^2+y^2+z^2\leq1.
	\end{equation}
	There exists the other parametrization called the probabilistic parametrization of qubit states \cite{Chernega1,Chernega2,Chernega3,Chernega4,Chernega5,Chernega6}. The matrix elements of the density matrix $\rho$ in this parametrization are expressed in terms of three nonnegative numbers $0\leq p_1,p_2,p_3\leq1$. These numbers satisfy the inequality
	\begin{equation}\label{eq.1.4}
	(p_1-1/2)^2+(p_2-1/2)^2+(p_3-1/2)^2\leq1/4.
	\end{equation}
	The Bloch ball parameters determine the points in the ball of the radius 1 with the center located in the point $x_0=y_0=z_0=0$. The probability parameters determine the points in another ball of the radius $1/2$ with the center located in the point $p_1^{(0)}=p_2^{(0)}=p_3^{(0)}=1/2$. The physical meaning of the Bloch parameters $x,y,z$ and the probability parameters $p_1,p_2,p_3$ are different. The numbers $p_j,\,j=1,2,3$ are the probabilities to have in the spin-1/2 state with density matrix $\rho$ the spin projections $m=+1/2$ on the three perpendicular directions determined by unit vectors $\vec e_1$, $\vec e_2$, $\vec e_3$ such that $(\vec e_i,\vec e_j)=\delta_{i j}.$ The Bloch parameters $x,y$ and $z$ are equal to mean values of the spin-$1/2$ projections on these directions. In view of this one has the relations between these parameters of the form 
	\begin{equation}\label{eq.1.5}
	x=2p_1-1,\,y=2p_2-1,\,z=2p_3-1.
	\end{equation}
	Geometrically the point in the Bloch ball can be connected with three angles $\theta_1,\theta_2$ and $\theta_3$ defined by scalar products of the vector $\vec r=(x,y,z)$ and the vectors $\vec e_j$, i.e. 
	\begin{equation}\label{eq.1.6}
	\cos\theta_j=(\vec r,\vec e_j)/|\vec r|.
	\end{equation}
	The three angle parameters satisfy the inequality 
	\begin{equation}\label{q.1.7}
	|\vec r|^2\left(\cos^2\theta_1+\cos^2\theta_2+\cos^2\theta_3\right)\leq1.
	\end{equation} 
	The probability representation of the spin-$1/2$ density matrix provides another trigonometric parameters for the point in the ball with radius $1/2$, with the center located in point $p_1^{(0)}=p_2^{(0)}=p_3^{(0)}=1/2$. One can introduce two angles $\phi$ and $\theta$ such that
	\begin{equation}\label{eq.1.8}
	\cos\phi=\frac{p_1-1/2}{\sqrt{(p_1-1/2)^2+(p_2-1/2)^2}},\quad 
	\sin\phi=\frac{p_2-1/2}{\sqrt{(p_1-1/2)^2+(p_2-1/2)^2}}.
	\end{equation}
	Another angle $\theta$ is determined by the probabilities $p_1,p_2$ and $p_3$ as 
	\begin{equation}\label{eq.1.9}
	\cos\theta=\frac{p_3-1/2}{\sqrt{(p_1-1/2)^2+(p_2-1/2)^2+(p_3-1/2)^2}}.
	\end{equation}
	For pure states with density matrix $\rho$ satisfying the purity condition $\mbox{Tr}\rho^2=1$ the number of parameters determining the density matrix equals to $2$. The angles $\theta_j$ determining the point of the Bloch sphere surface satisfy the equality 
	\begin{equation}\label{q.1.10}
	\cos^2\theta_1+\cos^2\theta_2+\cos^2\theta_3=1.
	\end{equation} 
	For pure states, the probability parameters $p_1,p_2,p_3$ satisfy the equality 
	\begin{equation}\label{q.1.11}
	(p_1-1/2)^2+(p_2-1/2)^2+(p_3-1/2)^2=1/4.
	\end{equation}
	One can see that for the pure spin-$1/2$ states one has relation of the angle $\theta$ with Bloch ball parameter $z$ given by formula 
	\begin{equation}\label{eq.1.12}
	\cos\theta=2z.
	\end{equation}
	
	\section{Photon states in probability representation}
	Let us consider one--mode photon states. There are different kinds of these states like Fock states $|n>$, $n=0,1,2,\ldots$ coherent states $|\alpha>$, where  $\alpha$ is a complex number, squeezed and correlated states \cite{DodKurmPLA} associated with the inequality determined by Schr\"odinger--Robertson uncertainty relation. In the conventional formulation of quantum mechanics, these states are described by the wave functions corresponding to the wave functions of harmonic oscillator with the Hamiltonian $\hat H=\hbar\omega(\hat a^\dag\hat a-1/2)$, where the photon creation $\hat a^\dag$ and annihilation $\hat a$ operators satisfy the bosonic commutation relation $[\hat a,\hat a^\dag]=1$. The photon number operator $\hat n=\hat a^\dag\hat a$ determines the Fock state $|n\rangle$ by the relation $\hat n|n\rangle=n|n\rangle$, $n=0,1,2,...$. The coherent state is the eigenstate of the non-hermitian annihilation operator $\hat a$, i.e., $\hat a|\alpha\rangle=\alpha|\alpha\rangle$. In the position representation, the wave functions $\langle x|n\rangle=\psi_n(x)$ of the Fock states reads 
	\begin{equation}\label{phst1}
	\psi_n(x)=\frac{\exp(-x^2/2)H_n(x)}{\pi^{1/4}\sqrt{2^nn!}},
	\quad(\hbar=m=\omega=1), 
	\end{equation}
	where $H_n(x)$ is Hermite polynomial. The coherent state wave function $\langle x|\alpha\rangle=\psi_\alpha(x)$ has the Gaussian form
	\begin{equation}\label{phst2}
	\psi_\alpha(x)=\frac{1}{\pi^{1/4}}\exp\left[-\frac{x^2}{2}-\frac{|\alpha|^2}{2}+\sqrt2\alpha x-\frac{\alpha^2}{2}\right],
	\end{equation}
	these wave functions are given as the superpositions of Fock states
	\begin{equation}\label{phst3}
	\psi_\alpha(x)=\exp(-\frac{|\alpha|^2}{2})\sum_{n=0}^\infty\frac{\alpha^n}{(n!)^{1/2}}\psi_n(x), 
	\end{equation}
	that corresponds to the connection of the state vectors $|n\rangle$ and $|\alpha\rangle $ of the form
	\begin{equation}\label{phst4}
	|\alpha\rangle=\exp(\alpha\hat a^\dag-\alpha^\ast\hat a)|0\rangle.
	\end{equation}
	The dimensionless dispersions of the position and momentum in coherent states are equal to $1/2$ and the coherent states are considered as maximally classical states of the photons. The normalized squeezed and correlated states of the photons have the normalized wave functions of generic Gaussian form 
	\begin{equation}\label{phst5}
	\psi_G(x)=\exp(-A x^2+B x+C),
	\end{equation}
	where the complex numbers $A$, $B$ and $C$ provide the equality
	\begin{equation}\label{phst6}
	(\delta x)^2(\delta p)^2=\frac{1}{4}\frac{1}{1-r^2},
	\end{equation}
	with $r$, the correlation coefficient of position and momentum. Under this condition the right-hand side of the equality is the bound in the Schr\"odinger--Robertson uncertainty relation, where $r$ is the correlation coefficient of the random position and momentum. In the probability representation of the photon states, these states are identified with symplectic tomographic probability distributions of photon quadrature $-\infty\leq X\leq\infty$, given by the formula \cite{MendesPLA} for arbitrary normalized wave function $\psi(x)$ of the form
	\begin{equation}\label{phst7}
	w(X|\mu,\nu)=\frac{1}{2\pi|\nu|}\left|\int\psi(y)\exp\left(\frac{i\mu y^2}{2\nu}-\frac{i X y}{\nu}\right)d y\right|^2.
	\end{equation} 
	Here $-\infty\leq\mu,\nu\leq\infty$ are parameters describing the reference frames in phase-space where the quadrature $X$ (a photon analog of the oscillator position $x$) is measured.
	
	The density operator $\hat\rho$ of the photon state is determined by the symplectic tomographic probability distribution by the relation (see, e.g. review \cite{IbortMarmoPhysSCR150})
	\begin{equation}\label{phst8}
	\hat\rho=\frac{1}{2\pi} \int w(X|\mu,\nu)\exp(i X-\mu\hat q-\nu\hat p)d Xd\mu d\nu.
	\end{equation}
	The density operator $\hat\rho$ determines the symplectic tomographic probability distribution \cite{ManciniTombesiPLA}
	\begin{equation}\label{phst9}
	w(X|\mu,\nu)=\mbox{Tr}\hat\rho\delta(X-\mu\hat q-\nu\hat p),
	\end{equation}
	where $\hat q$ and $\hat p$ are the position and momentum operators, respectively. For Fock states, the symplectic tomographic probability distributions are expressed in terms of Hermite polynomials
	\begin{equation}\label{phst10} 
	w(X|\mu,\nu)=\frac{\exp(-X^2)}{\sqrt{\pi(\mu^2+\nu^2)}}\frac{1}{2^nn!}H_n^2\left(\frac{X}{\sqrt{\mu^2+\nu^2}}\right).
	\end{equation}
	For coherent states, the tomographic probability distribution has the Gaussian form 
	\begin{equation}\label{phst11}
	w_\alpha(X|\mu,\nu)=\frac{1}{\sqrt{2\pi\sigma}}\exp\left[-\frac{(X-\bar X)^2}{\sigma}\right]
	\end{equation}
	with
	\[\bar X=\mu\sqrt2\,\mbox{Re}\alpha+\nu\sqrt2\,\mbox{Im}\alpha,\quad\sigma=\frac{\mu^2+\nu^2}{2}.
	\]
	For squeezed and correlated states with wave function $\psi_s(x)$ which has the form (\ref{phst5}) the tomographic probability distribution is determined by (\ref{phst11}) with the parameters $\bar X $ and $\sigma$ expressed in terms of the parameters $\mu$ and $\nu$ and $A,B,$ and $C$ of the form 
	\begin{equation}\label{phstK1}
	\sigma=\frac{|2A\nu-i\mu|^2}{2(A+A^\ast)},
	\end{equation}
	\begin{equation}\label{phstK2}
	\bar X=\frac{1}{2(A+A^\ast)}\left[\mu(B+B^\ast)+i\nu(2A B^\ast-2A^\ast B)\right].
	\end{equation}
	Thus, squeezed and correlated pure states of the photons are determined by normal probability distributions with dispersions of the photon quadratures (\ref{phstK1}) given by parameters $\sigma$, mean value (\ref{phstK2}) of the quadrature given by $\bar X$ and the covariance of the quadrature determined by the coefficient in front of term $\mu\nu$ in $\sigma$ (\ref{phstK1}). All the tomographic probability distributions of the photon states $|n>$, $|\alpha>$ and squeezed and correlated ones are normalized. They satisfy the condition
	\[\int w(X|\mu,\nu)d X=1\]
	for arbitrary parameters $\mu$, $\nu$. 
	The symplectic tomograms of the coherent states of the photon--phonon mode in Brillouin scattering process of light and in stimulated Raman scattering process, as well as  entanglement problem, were investigated in \cite{TchJRLR2001,JoptB2003}. The photon-number tomograms  and symplectic tomograms of photon states were considered in \cite{LaserPhys2009,Fortshchderphys2009,PhysScrip2010}. The Josephson junction devices and the model of resonant circuits were investigated in \cite{PhysScr2013,KiktenkoPhysRev,Kiktenkoar}. The symplectic tomogram of gaussian states of damped oscillator with $\delta$ -- kicked frequency was considered in  
	\cite{cherJLasRes2018}. 
Entropy and information associated with the probability representation of quantum states were considered in \cite{JRLR2007,Indus}.  
		
	\section{Coin probability distribution determining the photon states}
	One can extend the qubit states probability representation given by the $2\times2$ matrix $\rho(1)$ of the form 
	\begin{equation}\label{coinQ1}
	\rho=\left(\begin{array}{cc}
	p_3&(p_1-1/2)-i(p_2-1/2)\\
	p_1-1/2+i(p_2-1/2)&1-p_3
	\end{array}\right)
	\end{equation}
	writing it for qudit density $N\times N$-matrix $\rho_{j k}$ in terms of probability distributions of dichotomic random variables, e.g. of the form
	\begin{eqnarray}
	&&\rho_{j k}=p_1^{j k}-1/2-i(p_2^{j k}-1/2),\quad j>k,\nonumber\\
	&&\rho_{j j}-p_3^{j j},\quad j=1,2,\ldots,N-1,\nonumber\\
	&&\rho_{N N}=1-\sum_{j=1}^{N-1}p_3^{j j}.\label{coinQ2}
	\end{eqnarray} 
	For $N=\infty$ these relations are valid for arbitrary system where $\rho_{\infty\infty}$ is the limit in Eq.(\ref{coinQ2}) equal to zero. For example, the photon state density matrix given in the Fock basis $<n|\hat\rho|n'>$ can be expressed in terms of the probability distributions of random coin (dichotomic) variables where one has matrix elements of the density operator
	\begin{equation}\label{coinQ3}
	<n|\hat\rho|n'>=p_1^{(n, n')}-1/2-i(p_2^{(n, n')}-1/2),\, n<n',\,
	<n|\hat\rho|n>=p_3^{(n, n)},\,n,n'=0,1,2,\ldots.
	\end{equation}
	Here the probabilities $0\leq p_{1,2,3}^{(n,n')}\leq1$ satisfy the Silvester criterion. On the other hand, they are connected with the density matrix in position representation of the symplectic tomographic probability distribution $w(X|\mu,\nu)$ given by the equalities 
	\[\int<n|x><x|\hat\rho|x'><x'|n'>d x d x'=<n|\hat\rho|n'>\]
	or
	\begin{equation}\label{coinQ4}
	\frac{1}{2\pi}\int w(X|\mu,\nu)<n|e^{i(X-\mu\hat q-\nu\hat p)}|n'>d X d\mu d\nu=<n|\hat\rho|n'>,
	\end{equation}
	respectively. Here, the matrix elements of the Weyl system operator are given by analogs of formulas for the Franck--Condon factors of the harmonic oscillator
	\begin{equation}\label{coinQ5}
	<n|	e^{i(X-\mu\hat q-\nu\hat p)}|n'>=e^{i X}\int\psi_n(x)\psi_{n'}(x+\nu)e^{-i\mu x}d x.
	\end{equation}
	We calculate the integral (\ref{coinQ5}) using generating function method since the function
	\begin{equation}\label{coinQ6}
	f(\alpha,\beta^\ast)=\int\psi_\beta^\ast(x)\psi_\alpha(x+\nu)e^{-i\mu x}d x 
	\end{equation}
	determines the integral (\ref{coinQ5}), in view of the formula for coherent states and Fock states of the form
	\begin{equation}\label{coinQ7}
	f(\alpha,\beta^\ast)=\exp\left(\frac{-|\alpha|^2}{2}-\frac{|\beta|^2}{2}\right) \sum_{n,n'=0}^{\infty}\frac{\alpha^{n'}\beta^{\ast n}}{\sqrt{n!}}\int\psi_{n'}(x+\nu)\psi_n(x)e^{-i\mu x}d x.
	\end{equation}	
	The result can be expressed in terms of Hermite polynomials of two variables. Thus, one has different probability representations of photon states. One is the symplectic tomographic probability distribution $w(X|\mu,\nu)$. The other one is the infinite set of probability distributions of dichotomic random variables. The coin probability distribution and its time evolution is determined by the kinetic equation of the form of von Neumann equation for the matrix elements of the density matrix $\rho_{n n'}(t)$ rewritten for the probability distributions $p_{1,2,3}^{(j k)}(t)$. These distributions satisfy the kinetic equation
	\begin{equation}\label{eq.w0}
i\frac{\partial}{\partial t}\left(\rho_{j k}(t)\right)=
	\sum_{s=0}^\infty H_{j s}(t)\rho_{s k}(t)-\sum\rho_{j s}(t)H_{s k}(t),
	\end{equation}
	where the matrix $\rho_{s k}(t)$ is expressed by Eq.(\ref{coinQ3}) in terms of the probabilities and the Hamiltonian matrix elements are determined by the electromagnetic field conditions. For Fock states $|n\rangle$, the density matrix elements read $\rho_{n n'}=\delta_{n n'}$. This means that the probabilities determining the Fock states are 
	\begin{equation}\label{eq.w0'}
p_3^{(n n')}=\delta_{n n'}, \quad p_{1,2}^{(n n')}=\frac{1}{2}.
\end{equation}
	Thus, all the "coins" providing the Fock states are "ideal" for $n\neq n'$. It means that in such a game as coin flipping, coin tossing, or heads (UP) or tails (DOWN), which is the practice of throwing a coin in the air and checking which side is showing when it lands, in order to choose between two alternatives $p_k$ or $(1-p_k)$; $k=1,2,3.$  One of the coins corresponds to the probability $p_3^{(n n)}=1$. But for state $|n\rangle$ the $n$th coin is maximally nonideal always providing the position of "head" (or of the "tale"). For coherent state $|\alpha\rangle$ with matrix elements of the density matrix
	\begin{equation}\label{W1}
	<n|\alpha><\alpha|n'>=e^{-|\alpha|^2}\frac{\alpha^n\alpha^{\ast n'}}{\sqrt{n!n'!}}
	\end{equation}
	the coin probabilities $p_3^{(n n)}$ is the Poissonian 
	\begin{equation}\label{W2}
	p_3^{(n n)}=\frac{(\bar n)^n e^{-\bar n}}{n!},\quad \bar n=|\alpha|^2.
     \end{equation}
	The probabilities $p_1^{(n n')}$ and $p_2^{(n n')}$ are determined by the relation 
	\begin{equation}\label{W3}
	\frac{\alpha^n\alpha^{\ast n'}}{\sqrt{n!n'!}}e^{-|\alpha|^2}=p_1^{(n n')}-\frac{1}{2}-i(p_2^{(n n')}-\frac{1}{2}),\quad n<n'=0,1,2,\ldots.
	\end{equation}
	
	\section{The state evolution and energy levels in the probability representation of quantum mechanics}
In the probability representation of quantum mechanics the states are described by the sets of probability distributions. Any probability distribution of one random variable $0\leq p_1,p_2,\ldots p_N\leq1$ can be considered as a set of the probability distributions $(\pi_k,1-\pi_k),k=1,2,\ldots,N$ of $N$ dihotomic random variables, where $\pi_k=p_k$. For example, the Poissonian distribution of one random variable
\[p_1=\bar x e^{-\bar x},\quad p_2=\frac{\bar x^2}{2!}e^{-\bar x}, \quad p_3=\frac{\bar x^3}{3!}e^{-\bar x},\ldots\]
can be mapped onto the set of probability distributions $(\pi_1^{(1)}=\bar x e^{-\bar x},\pi_2^{(1)}=1-\bar x e^{-\bar x})$ and $ (\pi_1^{(2)}=\frac{\bar x^2}{2!}e^{-\bar x},\pi_2^{(2)}=1-\frac{\bar x^2}{2!}e^{-\bar x},\ldots$) of dihotomic random variables. Also a set of $N$ probability distributions of dichotomic random variables can be mapped onto the probability distribution of one random variable, using Bayes formula relating conditional probability distributions with the joint probability distribution. We demonstrate this map on example of qubit. Since the density matrix of qubit state is mapped onto three probability distributions of dichotomic random variables $(p_1,1-p_1),\,(p_2,1-p_2),\,(p_3,1-p_3)$ where $p_1,p_2,p_3$ are probabilities to get spin projections $m=+1/2$ onto three perpendicular directions, one can introduce the probability 6-vector $\vec\Pi=\frac{1}{3}\left(\Pi_1,\Pi_2,\ldots,\Pi_6\right)$ , where $\Pi_1=p_1, \Pi_2=1-p_1,\Pi_3=p_2,\Pi_4=1-p_2,\Pi_5=p_3,\Pi_6=1-p_3$.  
Thus, all the information on the qubit state density matrix is described by the probability distribution of two random variables $W(1,1)=\Pi_1,\,W(2,1)=\Pi_2,\,W(1,2)=\Pi_3,\,W(2,2)=\Pi_4,\,W(1,3)=\Pi_5,\,W(2,3)=\Pi_6$. Formally we introduce the probability distribution $W(j,k)$ of two random variables, such that ${\cal P}(k)=1/3$, $k=1,2,3$ is the marginal probability distribution, i.e. ${\cal P}(k)=\sum_{j=1}^2W(j,k)$ of the second random variable. Also, the conditional probability distributions $W(j|k)=W(j,k){\cal P}(k)$ provide the probability distributions of dichotomic random variables  $(p_1,1-p_1),\,(p_2,1-p_2),\,(p_3,1-p_3)$. The probability vector $\vec\Pi$ satisfies the linear evolution equation of the form 
\[\frac{d}{d t}\vec\Pi(t)=M \vec\Pi+\vec \gamma,\]
where $6\times 6$-matrix $M$ and vector $\vec \gamma$ are determined by the Hamiltonian matrix of the system. The stationary solution of the evolution equation, i. e. $\frac{d\vec\Pi}{d t}=0$ yields the energy spectrum of the system. The probability vector $\vec\Pi$ describing the states with the given energy levels does not depend on time. For harmonic oscillator with the Hamiltonian $\hat H=(\hat a^\dag\hat a+1/2)$ the evolution of coherent states of the complex parameter $\alpha=|\alpha|e^{i\phi_\alpha}$ reads 
$\alpha(t)=\alpha e^{-it}$. It means that the evolution of coherent states of photons is described by the evolution of probabilities 
\begin{equation}\label{E1}
p_1^{(n n')}(t)=\frac{1}{2}+\frac{|\alpha|^{n+n'}}{\sqrt{n!n'!}}\cos\left[(\phi_\alpha-t)(n+n')\right],
\end{equation}				
\begin{equation}\label{E2}
p_2^{(n n')}(t)=\frac{1}{2}-\frac{|\alpha|^{n+n'}}{\sqrt{n!n'!}}\sin\left[(\phi_\alpha-t)(n+n')\right].
\end{equation}			
The probability $p_3^{(n n')}(t)$ is the integral of motion for the coherent state evolution 
\begin{equation}\label{E3}
P_3^{(n n)}(t)=p_3^{(n n)}.
\end{equation}
The probabilities (\ref{E1})-(\ref{E2}) satisfy the kinetic equation (\ref{eq.w0}) for harmonic oscillator with matrix elements of the Hamiltonian $H_{j k}(t)=(\frac{1}{2}+j)\delta_{j k})$, where $j,k=0,1,2,\ldots,\infty.$ Since the numbers $p_{1,2,3}^{(n n')}(t)$ and the numbers $1-p_{1,2,3}^{(n n')}(t)$ determine the  probability distributions of dichotomic random variables, they satisfy the known entropic inequalities, e.g., relative entropy nonnegativity 
\begin{equation}\label{E4}
p_1^{(n n')}(t)\ln \left(\frac{p_2^{(n n')}(t)}{p_1^{(n n')}(t)}\right)+(1-p_1^{(n n')}(t))\ln \left(\frac{1-p_1^{(n n')}(t)}{1-p_2^{(n n')}(t)}\right)\geq0.
\end{equation}
In (\ref{E4}) the probabilities determining the time-dependent coherent state of harmonic oscillator are given by Eqs. (\ref{E1}) and (\ref{E2}). For squeezed and correlated state an analog of vacuum photon state can be described by the parametric oscillator with the Hamiltonian $\hat H(t)=\frac{\hat p^2}{2}+\frac{\omega^2(t)\hat q^2}{2}$, where $\omega(0)=1,\,\hbar=m=1$. The wave function of this state in position representation reads 
\begin{equation}\label{E5}
\psi_0(x,t)=\frac{1}{(\pi)^{1/4}}\frac{1}{\sqrt{\epsilon(t)}}\exp\left(\frac{i\dot\epsilon(t)}{2\epsilon(t)}x^2\right).
\end{equation}
The complex function $\epsilon(t)$ is the solution of the classical oscillator equation of motion 
\begin{equation}\label{E6}
\ddot\epsilon(t)+\omega^2(t)\epsilon(t)=0,
\end{equation}
with the initial conditions $\epsilon(0)=1$ and  $\dot\epsilon(0)=i$. The density matrix of the state $|0,t><0,t|$ in the Fock basis has the matrix elements 
\begin{equation}\label{E7}
\rho_{n n'}(t)=\int_{\infty}^{\infty}d x' d x\frac{H_n(x)H_{n'}(x')}{\sqrt{2^{n+n'}n!n'!\pi\epsilon(t)}}\exp\left(-x^2\left(\frac{1}{2}-\frac{i\dot\epsilon(t)}{2\epsilon(t)}\right)-\frac{x'^2}{2}\right).
\end{equation}
The probabilities $p_{1,2,3}^{(n n')}(t)$ determining the photon squeezed and correlated states given by Eq.(\ref{E5}) are expressed in terms of the integral (\ref{E7}).

\section{Transition probabilities in terms of tomograms}
In the conventional formulation of quantum mechanics the transition probabilities between the different states with wave functions $\psi_1(x)=\langle x|\psi_1\rangle$ and $\psi_2(x)=\langle x|\psi_2\rangle$ are given by the Born rule
\begin{equation}\label{B1}
p_{(1)}^{(2)}=|\int\psi_1^\ast(x)\psi_2(x) d x|^2.
\end{equation}
Here, the probability $p_{(1)}^{(2)}$ is expressed in terms of the wave functions identified with the quantum states. This transition probability can be expressed in terms of probabilities determining the system quantum states. For example, if one has the tomographic probability distribution $w_1(X|\mu,\nu)$ and $w_2(Y|\mu,\nu)$ determining the same states with state vectors $|\psi_1\rangle$  and $|\psi_2\rangle$,  the transition probability $p_{(1)}^{(2)}$ is expressed in terms of these tomographic probability distributions as follows 
\begin{equation}\label{B2}
p_{(1)}^{(2)}=\frac{1}{2\pi}\int w_1(X|\mu,\nu)w_2(Y|-\mu,-\nu)e^{i(X+Y)}d X d Y d\mu d\nu.
\end{equation}
In the molecular--spectroscopy theory, the expression given by Eq.(\ref{B1}) is called Frank--Condon factor. Thus, the Franck--Condon factor for vibronic structure of electronic lines of the polyatomic molecules is presented in the probability representation of quantum mechanics in the form of integral (\ref{B2}), where the integral kernel $e^{i(X+Y)}=\cos(X+Y)+i \sin(X+Y)$ can also be associated with the probability distributions of dichotomic random variables analogous to the distributions determining the qubit states, i. e.,
 \[\cos(X+Y)=\frac{(p_1-1/2)}{\sqrt{p_3(1-p_3)}}.\]
This example demonstrates that such quantum properties as transition probabilities between different states of a  system can be complitely described by the expressions containing standard probability distributions identified with these states.

For example, the transition probabilities between the squeezed and correlated states are expressed in terms of the integral
\begin{equation}\label{B3}
p_{(A_1,B_1,C_1)}^{(A_2,B_2,C_2)}=\frac{1}{4\pi^2}\int\frac{1}{\sqrt{\sigma_1\sigma_2}}\exp\left[-\frac{(X-X_1)^2}{2\sigma_1}-\frac{(Y-X_2)^2}{2\sigma_2}+i(X+Y)\right]d X d Y d\mu d\nu,
\end{equation} 
where parameters $\sigma_1$, $\sigma_2$, $\bar X_1$, and $\bar X_2$ are given by expressions 
(\ref{phstK1}) and (\ref{phstK2}), being dependent on the corresponding wave function parameters $A=A_1, B=B_1,$ and $ C=C_1$ for the first state and $A=A_2, B=B_2,$ and $ C=C_2$ for the second state. It is interesting that the overlap of classical normal distributions of random variables $X$ and $Y$ (\ref{B3}) with the integral kernel $\exp[i(X+Y)]$ provides the probability, which characterizes the quantum transitions between the squeezed and correlated states of a system. The evolution of pure qubit states can be described by the matrix $\rho(t_1,t_2)=|\psi(t_1
)\rangle\langle \psi(t_2)|$ expressed in terms of probability distributions $p_{1,2,3}(t_j)$, and  $1-p_{1,2,3}(t_j)$;  $j=1,2$ of dichotomic random variables of the form
\[
\rho(t_1,t_2)=\left(\begin{array}{cc}
\sqrt{p_3(t_1)p_3(t_2)}&
\sqrt{p_3(t_1)}\frac{[p_1(t_2)-1/2-i(p_2(t_2)-1/2)]}{\sqrt{p_3(t_2)}}\\
\sqrt{p_3(t_2)}\frac{[p_1(t_2)-1/2+i(p_2(t_2)-1/2)]}{\sqrt{p_3(t_1)}}&
\frac{[p_1(t_1)-1/2+i(p_2(t_1)-1/2)][p_1(t_2)-1/2-i(p_2(t_2)-1/2)]}{\sqrt{p_3(t_1)p_3(t_2)}}	
	\end{array}\right).
\]
In the case of $t_1=t_2=t$ the matrix $\rho(t_1,t_2)$ is equal to the qubit--state density matrix of qubit state in probability representation.

\section{Conclusion}
To conclude, we list the main results of our work. On examples, of photon mode (oscillator states) and spin-$1/2$ states (qubits), we demonstrated that the quantum mechanics formalism can be completely presented using the probability distributions of classical--like random variables. For photon states, one can use the quantum tomography approach where the system states are identified with probability densities of photon quadratures. For qubit (spin--$1/2$) systems, the states can be identified with probability distributions of dichotomic random variables. These results are valid for all other quantum system states. In the probability representation of quantum mechanics the system state density operators (oscillator, spins, hydrogen atom, molecules) and the matrix elements of density matrices are expressed in terms of probability distributions. The probability distributions are given by a vector $\vec{\cal P}(t)$ satisfying the linear kinetic equations of the form 
\[\frac{d}{d t}\vec{\cal P}(t)=M\vec{\cal P}(t)+\vec \gamma,\] where the matrix $M$ and vector $\vec \gamma$ are expressed in terms of matrix elements of the system Hamiltonian. The energy spectrum of the systems for time--independent Hamiltonians is determined by the condition $\frac{d}{d t}\vec{\cal P}(t)=0$. There exist an invertable map of probability vectors $\vec{\cal P}(t)$ onto the Hilbert space of states $|\psi(t)>$ and density operators $\hat\rho(t)$ acting in these spaces. Different kinds of probability vectors connected by linear transforms can be associated with the quantum states of the same system. We demonstrated that, for coherent states of the harmonic oscillator,  one can use symplectic tomographic probability distribution, where the state is identified with symplectic tomogram, which is normal distribution. Also the state can be identified with the set of probability distributions of dichotomic random variables. Both descriptions are related by means of linear transforms of the distributions. We also constructed the probability distribution of the squeezed vacuum photon state and obtained new entropic inequalities for the matrix elements of the parametric oscillator density matrix. The consideration of other aspects of the probability representation of quantum mechanics and applying the probabilistic approach to studying quantum technologies will be done in future publications.

\section*{Acknowledgments}
This work was supported by the Russian Science Foundation under
Grant No. 19-71-10091. V.  I. Man'ko thanks the Organizers of 26th Central European Workshop on Quantum Optics (Paderborn Universityh, Germany, June 3-7, 2019) for the invitation.

\end{document}